\documentstyle[12pt,aaspp4]{article}

\def\degree{\ifmmode {^\circ}\else {$^\circ$}\fi}
\def\mum{\ifmmode {\rm \mu {\rm m}}\else $\rm \mu {\rm m}$\fi}
\def\arcsec{\ifmmode ^{\prime \prime}\else $^{\prime \prime}$\fi}
\def\secpoint{\mbox{$''\mskip-7.6mu.\,$}}

\def\inch{\ifmmode ^{\prime \prime}\else $^{\prime \prime}$\fi}
\def\arcmin{\ifmmode ^{\prime}\else $^{\prime}$\fi}

\def\lsun{\ifmmode {\rm L_{\odot}}\else $\rm L_{\odot}$\fi}
\def\msun{\ifmmode {\rm M_{\odot}}\else $\rm M_{\odot}$\fi}
\def\msunyr{\ifmmode {\rm M_{\odot}~yr^{-1}}\else $\rm M_{\odot}~yr^{-1}$\fi}
\def\mearth{\ifmmode {\rm M_{+\mskip-14.6muO\,}}\else $\rm M_{+\mskip-14.6muO\,}$\fi}
\def\mearth{\ifmmode {\rm M_{\earth}}\else $\rm M_{\earth}$\fi}
\def\etal{{\it et al}. }

\def\singlespace{\baselineskip=14pt}

\begin{document}

%
%
\title{Optical Spectroscopy of Embedded Young Stars 
in the Taurus-Auriga Molecular Cloud}
\singlespace

\author{Scott J. Kenyon}
\affil{Harvard-Smithsonian Center for Astrophysics,
60 Garden Street, Cambridge, MA 02138} 
\affil{e-mail: skenyon@cfa.harvard.edu}

\author{David I. Brown}
\affil{Jet Propulsion Laboratory, MS 156-205, 
4800 Oak Grove Drive, Pasadena, CA 91109}
\affil{e-mail: dib@lb6.jpl.nasa.gov}

\author{Christopher A. Tout}
\affil{Institute of Astronomy, Madingley Road, Cambridge CB3 0HA England}
\affil{e-mail: cat@ast.cam.ac.uk}

\and

\author{Perry Berlind}
\affil{Whipple Observatory, PO Box 97, Amado, AZ 85645}
\affil{e-mail: pberlind@cfa.harvard.edu}

\vskip 4ex
\centerline{to appear in }
\centerline{\it The Astronomical Journal}
\centerline{June 1998}

\vskip 4ex
\received{5 December 1997}
\accepted{}

\clearpage

%
%

\singlespace

\begin{abstract}


This paper describes the first optical spectroscopic survey of
class I sources (also known as embedded sources and protostars) 
in the Taurus-Auriga dark cloud.  We detect 10 of the 24
known class I sources in the cloud at 5500--9000 \AA.  All
detected class I sources have strong H$\alpha$ emission;  
most also have strong [O~I] and [S~II] emission.  These data --
together with high quality optical spectra of T Tauri stars in
the Taurus-Auriga cloud --
demonstrate that forbidden emission lines are stronger and more 
common in class I sources than in T Tauri stars.  Our results 
also provide a clear discriminant in the frequency of forbidden
line emission between weak-emission and classical T Tauri stars.
In addition to strong emission lines, three class I sources have 
prominent TiO absorption bands.  The M-type central stars of these 
sources mingle with optically visible T Tauri stars in the 
HR diagram and lie somewhat below both the birthline for 
spherical accretion and the deuterium burning sequence for disc accretion.

\end{abstract}


%
%

\section{INTRODUCTION}

Examining the earliest phases of low mass stellar evolution requires 
observations of protostars deeply embedded in the dense cores of 
nearby molecular clouds.  These ``class I'' sources (\cite{lad87})
have blackbody-like spectral energy distributions that peak at 
wavelengths of 30--100 \mum~and bolometric luminosities,
$L_{\rm b} \approx$ 0.1--100 \lsun~(\cite{ada87}; \cite{mye87}; 
\cite{wil89}; \cite{ken90}; \cite{gre94}).  Despite many detailed 
studies of their circumstellar environments (see, for example,
\cite{ta91a},b; \cite{and93}; \cite{ter93}; \cite{mor92}, 1995; 
\cite{bon96}; \cite{gom97}; \cite{hog97}; \cite{whi97}), 
understanding the stellar physics of these objects has proved elusive.
Comparisons of observed bolometric luminosity functions with models
is straightforward but controversial (\cite{wil89}; \cite{ken90};
\cite{fl94a}, 1994b).  The apparent lack of photospheric features 
in many objects
has led several groups to abandon the HR diagram as a means for
testing stellar evolutionary tracks of the youngest stars.  
These groups have proposed the bolometric temperature 
(\cite{mye98} and references therein), 
the submillimeter flux (\cite{sar96}), and the visual extinction
(\cite{ada90}) to replace effective temperature and have 
developed models to place evolving pre--main-sequence stars in their 
modified evolutionary diagrams.  The accuracy of these techniques 
remains uncertain, because the methods are new and incompletely
tested.

In this paper, we report an optical spectroscopic survey designed
to detect photospheric absorption features from the central stars of 
class I sources in the Taurus-Auriga cloud.  Although the line-of-sight
extinction to the central star is large, $A_V \approx$ 30--60 mag
(\cite{whi97}), large ground-based telescopes can detect optical light 
scattered off cavities in the infalling envelopes of many objects.
Optical data also provide the best measure of spectral types for
pre--main-sequence stars.  In general, I-band and J-band data are 
least contaminated by emission from an accretion disc and its 
associated boundary layer or magnetic accretion column (\cite{kh90}).  
However, the very large continuum veiling detected on near-IR spectra 
of some class I sources (\cite{cas92}; \cite{gr96a}, 1996b) favors 
I-band spectra,
because the J-band veiling can be large if the disc extends to the
stellar photosphere (\cite{kh90}; \cite{ken96}).  Finally, optical 
spectra of class I sources allow an unambiguous comparison with 
optically brighter T Tauri stars, which have known spectral types 
in a well-calibrated system (see, for example, Kenyon \& Hartmann 1995;
KH95 hereafter).

Our results provide the first optical detection of M-type absorption features
in an embedded protostar.  We identify TiO absorption bands in three
Taurus-Auriga class I sources; one other star may have TiO features
and a fifth star may have K-type absorption features.  We use optical
spectra of T Tauri stars to calibrate the spectral types of class I
sources and then construct a complete HR diagram for the Taurus-Auriga 
cloud.  These data, coupled with new evolutionary tracks for protostars
accreting from discs and two spectral types derived from near-IR spectra
(\cite{gr96b}), show that class I sources in Taurus-Auriga mingle with 
T Tauri stars and lie below the birthline in the HR diagram.

We also detect strong emission lines on the spectra of all protostars.
Forbidden emission from [N~II] and [S~II] is much more common among
class I sources than older, optically brighter stars having the same
bolometric luminosity.  The fluxes of forbidden emission lines also
seem stronger among class I sources than other pre--main-sequence stars
in the cloud.  We find no evidence that the permitted emission lines,
such as H$\alpha$ and He~I, are more common or stronger than in 
T Tauri stars.  These results extend and confirm previous conclusions 
that jet activity declines as a pre--main-sequence star contracts to
the main-sequence.

We describe our observations in Sec. 2, explain our results in Sec. 3, 
and conclude with a brief discussion in Sec. 4.

%
%

\section{OBSERVATIONS}

We acquired optical spectra of faint Taurus-Auriga class I sources
and other pre--main-sequence stars with the Red Channel Spectrograph 
at the Multiple Mirror Telescope (MMT; 16--19 November 1995)
and the Double Spectrograph at the Palomar 5-m telescope
(29--30 November 1995).
At the MMT, we used the 270 g mm$^{-1}$ grating and a 
1\arcsec~slit to produce spectra covering 5700--9000 
\AA~on a 1200 $\times$ 800 Loral CCD.  
On-chip binning of the pixels, 2 $\times$ 2,
yielded a spectral resolution of 10.8 \AA~and 
a spatial resolution of 0\secpoint6 per pixel.
At Palomar, we used a 316 g mm$^{-1}$ grating, a
2\arcsec~slit, and a 1024 $\times$ 1024 CCD.  
The Palomar spectra cover 6000--8500 \AA~with
a spectral resolution of 10 \AA~and a spatial
resolution of 0\secpoint47~per pixel.

We obtained low resolution spectra of brighter young stars 
in Taurus-Auriga during 1995--1996 with FAST, a 
high throughput, slit spectrograph mounted at the 
Fred L. Whipple Observatory 1.5-m telescope on 
Mount Hopkins, Arizona (\cite{fab98}).
We used a 300 g mm$^{-1}$ grating  blazed at 4750 \AA, 
a 3\arcsec~slit, and recorded the spectra on a thinned Loral 
512 $\times$ 2688 CCD.  These spectra cover 3800--7500 \AA~at 
a resolution of $\sim$ 6 \AA.

We derive final object spectra using standard tasks
within NOAO IRAF.  After trimming the CCD frames at
each end of the slit, we correct for the bias level,
flat-field each frame, apply an illumination correction,
and derive a full wavelength solution from calibration
lamps acquired immediately after each exposure.  The 
wavelength solution for each frame has a probable error 
of $\pm$0.5--1.0 \AA.
We extract object and sky spectra using the optimal
extraction algorithm within APEXTRACT.  We vary the 
size of the object/sky aperture from source to source
to include additional radiation from extended emission.

At both the MMT and Palomar, we selected class I sources based 
on published VRI photometry (see KH95), POSS red plates, and 
red narrow band continuum images (\cite{gom97}).  Our sample 
of 10 class I sources is complete to R $\sim$ 20: we detected 
all sources with optical counterparts on the POSS and several 
sources with measured R $\sim$ 19--20.  We did not observe 14 
other class I sources without known optical counterparts and
cannot estimate a 
reliable detection frequency for deeply embedded sources.
The observed sample spans the observed range of class I luminosities
in the Taurus-Auriga cloud, $L_{\rm b} \sim$ 0.2--20 \lsun, 
but does not include the reddest systems that do not have 
optical counterparts (see \cite{whi97}).

We selected FAST sources from the KH95 sample of $\sim$ 150
known pre--main-sequence stars in the Taurus-Auriga cloud.
We observed essentially all targets with V $\lesssim$ 16--17:
55 out of 65 weak emission T Tauri stars,
69 out of 96 classical T Tauri stars,
and one class I source with a bright 
optical counterpart (Haro 6-28).
Aside from their apparent brightness, these samples do not
appear to be biased against any particular observational
property of pre--main-sequence stars: the distributions
of H$\alpha$ equivalent widths, K--L colors, and bolometric
luminosities of stars in the FAST sample are indistinguishable 
from the distributions for stars not included in the FAST sample
using data from Cohen \& Kuhi (1979) and KH95.
We thus conclude that the FAST objects are a representative sample
of known pre--main-sequence stars in the cloud.

\section{RESULTS}

\subsection{Basic Properties and Spectral Types}

Figure 1 shows FAST spectra of T Tauri stars with a range
of emission characteristics.  Weak emission T Tauri stars,
such as LkCa 3, have absorption spectra similar to normal
main-sequence stars with additional weak H$\alpha$ emission 
lines (\cite{coh79}; \cite{wal87}).
Classical T Tauri stars, such as BP Tau, DG Tau, and DP Tau, 
have prominent emission lines and a variable blue continuum 
superimposed on a late-type absorption spectrum (\cite{coh79};
\cite{har89}, 1991).  
Some T Tauri stars have few emission lines other than H~I 
and He~I; others have prominent [O~I], [S~II], and [Fe~II] 
emission lines (\cite{coh79}).  In most 
interpretations, the H~I and He~I lines form in 
an accretion region or outflowing wind close to 
the central star; the [O~I] and [S~II] lines form
in a jet or in the wind (\cite{bas93}; \cite{edw87}; 
\cite{ham92}; see also \cite{mar97}).

Figure 2 shows contour maps of MMT spectra for several sources 
centered on H$\alpha$ and [S~II] $\lambda\lambda$6717,6730.
We fit gaussian profiles to the continuum and a few emission
lines along the spatial direction for each class I source and 
three point sources.  Several class I sources -- such as 
L1489 IRS (04016+2610), HH31 IRS2 (04248+2612), and 04264+2433 -- 
are clearly extended along the slit, with $\sigma \approx$ 
5\arcsec--10\arcsec~compared to $\sigma \approx$ 
1\secpoint0$\pm$0\secpoint2 for point sources.  These sizes are
comparable to sizes inferred from optical and near-infrared
images of these sources (\cite{whi97}; \cite{gom97}, and 
references therein).  The very deeply embedded sources L1527 IRS
(04368+2557) and L1551 IRS5 (04287+1801) are much more extended than 
other class I sources.  Both have large optical reflection nebulae, 
$\gtrsim$ 30\arcsec~across, with multiple emission knots (see, for 
example, \cite{sto88}; \cite{gra92}; \cite{eir94}; \cite{gom97}). 

Our results indicate that the continuum and emission lines of all 
class I sources are equally extended within the errors of the fit.
The mean difference in spatial extent between the emission lines
and the continuum is $\langle \sigma \rangle$ = 0\secpoint4 $\pm$
0\secpoint3 for the 5 sources with strong continua (L1489 IRS, 
04158+2805, HH31 IRS2, 04264+2433, and 04489+3042).  This difference
is small compared to the typical spatial extent of each source,
$\sigma \approx$ 5\arcsec--10\arcsec.  Our observations have 
insufficient spatial resolution to discriminate between sources 
with emission knots such as L1527 IRS and those without emission knots
such as 04489+3042.

Figures 3--5 show MMT and Palomar spectra for Taurus class I sources.
We detect a strong continuum in several objects (Figure 3),
including L1489 IRS, HH31 IRS2, 
04264+2433, and 04489+3042.  Three class I sources -- 
04158+2805, 04248+2612 and 04489+3042 -- have the 
deep TiO absorption bands characteristic of M-type
stars.  The continua of L1489 IRS and 04264+2433 appear 
featureless, although both have a prominent dip at 8100 \AA.
All of these class I sources have a strong H$\alpha$ emission line,
along with moderately strong emission from [O~I], [S~II], and Ca II.
Several have He~I emission at $\lambda\lambda$5876, 6678, 7065.
These spectra are similar to the spectra of classical T Tauri 
stars in Figure 1.

The spectra of class I sources in Figure 5 more closely resemble 
spectra of jets or Herbig-Haro (HH) objects (e.g., \cite{boh90};
\cite{dop82}; \cite{goo86}; \cite{rei91}; see also \cite{rag96}).
We detect little, if any, continuum emission from 04239+2436,
04295+2251, and L1527 IRS, but all of these objects have very
strong emission lines (Figure 5).  The relative intensities
of the emission lines in 04239+2436 and L1527 IRS are similar
to those for on-source emission in HH30 IRS (Figure 4; right
panel) and L1551 IRS5 (Figure 5; upper right panel).
We detect [O~I] and [S~II] emission in all class I sources
except 04295+2251 (Figure 5; lower left panel), where 
we identify a prominent H$\alpha$ emission line superimposed
on a very weak continuum.

To compare spectra of class I sources and T Tauri stars in more
detail, we measure the strengths of several prominent absorption 
and emission lines.  We fit gaussian profiles to obvious emission
lines using SPLOT within NOAO IRAF and use the deblend option
for blended lines such as the [S~II] doublet and the He~I
$\lambda$5876 and Na~I blend.  In the absence of accurate
de-reddened fluxes, the equivalent width, EW, provides a good 
relative measure of emission line strengths for strong continuum
sources.  We place upper limits of 100--200 \AA~for equivalent
widths of weak continuum sources depending on the continuum
level.  Table 1 lists our results for class I sources; Table 2
summarizes measurements for FAST spectra of T Tauri stars.
We include separate entries for the MMT and Palomar spectra 
of two objects, L1489 IRS and HH31 IRS2, and list the MMT
results first in both cases.
We estimate probable errors of $\pm$10\% for strong lines
with EW $>$ 10 \AA~and $\pm$20\% for weaker lines based on
measurements of 2--3 separate spectra for each star in the 
FAST sample and two stars in the MMT/Palomar sample.

We measure spectral types using TiO absorption indices,
defined as the depth of a TiO band at a wavelength $\lambda$
relative to an interpolated continuum point (\cite{oco73}).
The TiO bands at $\lambda$6180 and $\lambda$7100 are 
temperature-sensitive for M dwarfs and giants (\cite{oco73}).
Kenyon \& Fern\'andez-Castro (1987) derive reliable spectral types 
for the red giant components in symbiotic stars -- which also have
strong emission lines and a variable blue continuum -- with
these features.  We define:

\begin{equation}
\rm [TiO]_1 = -2.5~log \left ( \frac{F_{6180}}{F_{6125} + 0.225(F_{6370}~-~F_{6125})} \right )
\end{equation}

\noindent
and

\begin{equation} 
\rm [TiO]_2 = -2.5~log \left ( \frac{F_{7100}}{F_{7025} + 0.2(F_{7400}~-~F_{7025})}  \right )  ~~ .
\end{equation}

\noindent
Each 30 \AA~bandpass used for these indices avoids contamination 
from strong emission lines and telluric absorption bands.  The two
TiO indices increase from [TiO]$_1$ $\approx$ [TiO]$_2$ $\approx$ 0
at K4--K5 spectral types to [TiO]$_1$ $\approx$ 0.8 and [TiO]$_2$ 
$\approx$ 1.0 at M6 spectral types.

Figure 6 shows [TiO]$_2$ as a function of [TiO]$_1$ for normal
main-sequence stars (filled circles), T Tauri stars with weak
emission lines on FAST spectra (light triangles), and pre--main
sequence stars with MMT or Palomar spectra (crosses and diamonds).
The locus of T Tauri stars generally follows the main-sequence 
stars except near [TiO]$_1$ $\approx$ 0.5, where pre--main-sequence
stars have larger [TiO]$_2$ indices compared to main-sequence stars.
Most pre--main-sequence stars with MMT or Palomar spectra lie on
the T Tauri star locus.  Both L1489 IRS and 04303+2240 have 
featureless continua and negligible TiO absorption on their spectra.
Other stars with strong optical continua have modest to strong TiO 
absorption bands and must have M-type central stars.  

The measured TiO indices indicate optical veiling in two sources with 
MMT or Palomar spectra.  The class I sources HH32 IRS2 and 04158+2805 
lie above the pre--main-sequence locus in Figure 6.
In both cases, the M-type absorption features for $\lambda >$ 7400 \AA~are
very strong, which suggests that the [TiO]$_1$ index is `weak' compared 
to the [TiO]$_2$ index.  Many T Tauri stars with strong emission lines 
also have weak [TiO]$_1$ indices.  These T Tauri stars have substantial
emission from a blue continuum source which veils optical absorption lines;
this veiling increases towards short wavelengths in all cases (see
\cite{har91}).  Optical veiling from a hot, $T \sim 10^4$ K, continuum 
source probably causes the weak [TiO]$_1$ index in HH32 IRS2 and 
04158+2805, but our optical spectra have insufficient signal-to-noise
to verify that absorption features at $\lambda <$ 6000 \AA~are similarly
weakened.

We derive spectral types for class I sources and faint T Tauri stars
via comparison with T Tauri stars of known spectral type.
We adopt spectral types for bright T Tauri stars from KH95 and use 
FAST spectra to calibrate
[TiO]$_1$ and [TiO]$_2$ as a function of spectral type.
The measured TiO indices for class I sources then yield the 
spectral types listed in Table 1.  We estimate probable 
errors of $\pm$1--2 subclasses for the spectral types based
on measurement errors of the TiO indices and the intrinsic
uncertainty in assigning spectral types to bright T Tauri stars.

\subsection{Jet Emission}

As we noted in the introduction, powerful optical and molecular
outflows distinguish class I sources from other pre--main-sequence
stars in nearby dark clouds.  Practically all class I sources have
molecular outflows; very few optically visible T Tauri stars are
associated with molecular outflows (e.g., \cite{bon96}). In their 
complete survey of Taurus class I sources, G\'omez \etal (1997) show 
that the frequency of optical jets decreases from $\gtrsim$ 60\% for
class I sources to $\lesssim$ 10\% for T Tauri stars.  Among T Tauri
stars, optical jet emission is almost always associated with classical 
T Tauri stars (CTTS) instead of weak-emission T Tauri stars (WTTS; see 
\cite{edw87}; \cite{har95}).  Near-infrared surveys also indicate 
more emission line activity among class I sources than CTTS or WTTS
(\cite{gr96a}, 1996b).  These results suggest that jet emission is
correlated with disc accretion and that disc accretion somehow declines
from class I sources to T Tauri stars (\cite{edw87}; \cite{har95};
\cite{gr96b}).

To examine these correlations with our spectroscopic data, we divide sources 
into classes based on the ratio of far-infrared to bolometric luminosity,
$L_{FIR}/L_{\rm b}$ (KH95).  In this system, class I sources have 
$L_{FIR}/L_{\rm b} \ge 0.8$,
CTTS have $L_{FIR}/L_{\rm b} \approx$ 0.1--0.3, 
and WTTS sources have $L_{FIR}/L_{\rm b} \le 0.1$.  
We also select 14 sources in Tables 1--2 with 
$L_{FIR}/L_{\rm b} \approx$ 0.3--0.8 as flat-spectrum
sources (see also KH95; \cite{gr96b}, 1997).
All of these sources have the same median luminosity, $L_{\rm b} 
\approx$ 0.5--0.8 \lsun, except for the flat-spectrum which have
$L_{\rm b} \approx$ 1.5 \lsun~(KH95).
For each class, we compute the detection frequency for each of the emission 
lines listed in Tables 1--2.  Figure 7 shows our results for [S~II].
The frequency of [S~II] emission obviously increases with 
increasing $L_{FIR}/L_{\rm b}$.  
We find a similarly striking trend for [N~II] emission: practically all 
class I sources have strong [N~II] but only a few CTTS or WTTS have 
any [N~II] emission.

Our detection frequencies for forbidden-line emission are lower limits,
because we cannot detect weak emission lines with EW $\lesssim$ 0.2--0.4 
\AA~(see Tables 1--2).  To estimate the importance of this uncertainty,
we compare our results with Hartigan \etal (1995) who derive emission
line equivalent widths from echelle spectra.  At higher resolution,
the detection frequency for [S~II] emission increases to $\sim$ 50\%
for CTTS and remains unchanged at 0\% for WTTS.  The [N~II] emission shows
a similar trend but is detected less often than [S~II]. Hartigan \etal 
note, however, that the forbidden emission lines in CTTS consist of 
high velocity material from the jet and low velocity gas near the disc.  
The weak, low velocity emission is responsible for the larger [S~II]
detection frequency among CTTS in the Hartigan \etal sample.  Our low 
resolution spectra do not detect this emission and thus provide a
proper estimate for the frequency of high velocity jet emission among CTTS.

To check further the reality of the trend in Figure 7, we perform a simple 
test.  We assume a parent population of $N$ sources with an intrinsic 
probability, $p$, of [S~II] emission.  For $n_j$ observed sources, 
the probability of detecting [S~II] emission in $k_j$ sources is given 
by the binomial distribution:

\begin{equation}
p_{obs} = \frac{n_j!}{k_j!~(n_j-k_j)!} ~ p^{k_j}~(1-p)^{(n_j-k_j)} ~ 
\end{equation}

\noindent
for $n_j < N$.
If we require $p_{obs} \gtrsim 10^{-3}$, the allowed range in $p$ 
for a single class of pre--main-sequence star is large:
$p \lesssim$ 0.13 for WTTS, $p \approx$ 0.03--0.36 for CTTS,
$p \approx$ 0.24--0.93 for flat-spectrum sources, and
$p \approx$ 0.33--1.00 for class I sources.
However, the probability of realizing the observed frequency of
[S~II] detection for any two classes from a single parent population 
is extremely small.   The probability that the WTTS and CTTS in our
sample have the same parent distribution never exceeds $10^{-3}$; 
it exceeds $10^{-4}$ only for $p$ = 0.04--0.12.  We find no common 
intrinsic probability for the CTTS and flat spectrum sources or 
for CTTS and class I sources: $p_{obs}$ is less than $10^{-4}$ for 
any value of $p$.  The observed frequencies of [S~II] emission 
in class I and flat spectrum sources, however, could be chosen from 
the same parent population for $p$ = 0.46--0.90 if $p_{obs} \ge 10^{-3} $.  

We derive similar results for the frequency of [N II] emission.
We detect the $\lambda\lambda$6548, 6584 doublet in
82\% of 11 class I sources, 36\% of 14 flat spectrum sources,
4\% of 46 CTTS, and 0\% of 54 WTTS with reliable $L_{FIR}/L_{\rm b}$.
The allowed ranges in the intrinsic probability of [N~II] emission
are then $p \gtrsim$ 0.33 for class I sources, 
$p \approx$ 0.07-0.76 for flat-spectrum sources,
$p \lesssim$ 0.21 for CTTS, and $p \lesssim$ 0.12 for WTTS.
The observed detection frequencies allow a single intrinsic probability
for [N~II] emission between CTTS and WTTS for 
$p \lesssim$ 0.11 and between class I and flat
spectrum sources for $p \approx$ 0.40--0.71.
Our data do not allow a single [N~II] emission probability for 
all sources.  This result is not as strong as for the [S~II] lines,
because weak [N~II] emission is more difficult to detect due to
the strong H$\alpha$ lines in many CTTS.

We conclude that the increasing frequency of forbidden line emission 
as a function of $L_{FIR}/L_{\rm b}$ is real.  Sources with large 
$L_{FIR}/L_{\rm b}$ are much more likely to be associated with 
forbidden line emission than sources with small $L_{FIR}/L_{\rm b}$.  
In addition, our sample is large enough to detect a significant 
difference in the frequency of [S~II] emission between CTTS and WTTS.
Previously published data had suggested this difference, but the data 
were too heterogeneous to make a firm statistical comparison.  
Our sample is {\it not} large enough to detect a difference in the [N~II] 
or [S~II] emission frequency between class I and flat spectrum sources.
Adding sources to the optical sample and improved measurements of 
$L_{FIR}/L_{\rm b}$ would allow a better discriminant between class I
and flat spectrum sources.  These improvements require larger 
ground-based telescopes and new far-IR data from either ISO or SIRTF.

\subsection{The HR Diagram}

Figure 8 shows an HR diagram for Taurus-Auriga pre--main-sequence stars 
(see KH95).   Filled circles plot WTTS and CTTS from KH95\footnote{We use
the KH95 conversion from spectral type to effective temperature for 
the class I sources.}.  Crosses indicate the positions of 04158+2805,
HH31 IRS2, and 04489+3042 using our new spectral types.  The triangle
indicates Haro 6-28 with a revised spectral type (M2) based on FAST
spectra.  The diamonds denote 04181+2655 and 04295+2251 using near-IR
spectra from Greene \& Lada (1996b).  The relative positions of these
two class I sources should be accepted with some caution, because 
near-IR spectral types for 04489+3042 (K1-K2) and Haro 6-28 (K5-K6) 
are much earlier than our optical spectral types.  

Although the class I sample is small, the positions of class I sources 
are not especially distinct from the distribution of WTTS and CTTS. 
Three class I sources straddle the $10^6$ yr isochrone from the CMA 
models of D'Antona \& Mazzitelli (1994).  Two other class I sources 
fall midway between the $10^5$ yr and $10^6$ yr isochrones.  All 
class I sources are within 2$\sigma$ of the $10^5$ yr isochrone 
and the stellar ``birthline'' for spherical accretion 
(\cite{sta83}, 1988; \cite{pal90}, 1993; see also KH95).
Only Haro 6-28, however, lies {\it on} the stellar birthline\footnote{We
suspect that Haro 6-28 may be a close binary similar to GV Tau (Haro 6-10),
because their spectral energy distributions are similar.  Both
GV Tau and Haro 6-28 are 2--3 mag brighter in the optical than
other class I sources in the sample.}

To compare our new data with an alternative to spherical protostellar 
accretion theory (\cite{sta83}, 1988), we consider the evolution 
of protostars accreting material from a disc.  We construct a set 
of stellar models using the most recent version of the Eggleton 
evolution program (\cite{egg71}, 1972, 1973).  Our models assume 
an initially uniform composition with abundances of 
hydrogen $X = 0.7$, helium $Y = 0.28$, 
deuterium $X_{\rm D} = 3.5\times 10^{-5}$, and 
metals $Z = 0.02$ appropriate for the meteoritic mixture 
determined by Anders \& Grevesse (1989).  
Pols \etal (1995) describe the equation of state, which includes 
molecular hydrogen, pressure ionization, and coulomb interactions.
The nuclear reaction network includes the pp chain and the CNO cycles.  
Deuterium burning is explicitly included at temperatures too low for the 
pp chain.  Once the pp chain is active hydrogen burns to He$^4$ with deuterium 
and He$^3$ in equilibrium.  The burning of He$^3$ is not explicitly followed.
We use the opacity tables of Iglesias, Rogers \& Wilson (1992) and Alexander
\& Ferguson (1994).  We adopt an Eddington approximation (\cite{woo53})
for the surface boundary conditions at an optical depth of $\tau = 2/3$.
Low-temperature atmospheres, in which convection extends out as far as 
$\tau \approx 0.01$ (\cite{bar95}), are not modeled completely.  
However the effect on observable quantities is not significant
(see \cite{kro98}).

In these calculations, the initial protostar is a fully convective 
$0.1\,M_\odot$ object with a radius of $3\,R_\odot$ and an effective 
temperature of $10^{3.43}$ K.  This starting point lies just off the 
right boundary of Figure~8.  We add accreted material to the surface 
with the initial composition and with the same state (entropy, temperature, 
etc.) as the surface.  In this approximation, most of the stellar surface 
is free to radiate normally with the boundary conditions described above.  
These boundary conditions are a compromise, because we are modeling a two 
dimensional process with a one dimensional evolution code.  

The solid lines in Figure~8 indicate two accreting protostellar tracks 
using the Eggleton code.  The thinner line accretes at 
$\rm 10^{-6}\,M_\odot\,{\rm yr}^{-1}$; the thick line accretes at 
$\rm 10^{-5}\,M_\odot\,{\rm yr}^{-1}$.  Initially, both stars contract at 
fairly constant luminosity and move to the left in the HR diagram.
This contraction continues until deuterium ignites at their centers. 
The tracks then turn upwards to follow a deuterium burning sequence. 
Both protostars are fully convective, so newly accreted material 
replenishes central deuterium.  These protostars thus remain on the 
deuterium burning track until either accretion ceases or the rate
of deuterium replenishment becomes insufficient for burning to continue 
to support the star.  The latter occurs first, temporally, for the
$\rm 10^{-5}\,M_\odot\,{\rm yr}^{-1}$  model at $T_{\rm eff} = 10^{3.62}\,$K
when it has reached a mass of $0.686\,M_\odot$. The
$\rm 10^{-6}\,M_\odot\,{\rm yr}^{-1}$ model drops below the deuterium sequence
at $T_{\rm eff} = 10^{3.58}\,$K at a mass of $0.456\,M_\odot$.
In both cases, we continue accretion until the total mass reaches
$0.96\,M_\odot$.  During the protostellar evolution, the stars lie
close to the Hayashi tracks appropriate to their instantaneous mass.
If accretion ceases at any time, the star will shrink down to the 
main-sequence along a normal pre--main-sequence track and will
ignite and burn deuterium on the way if it has not already done so.

In our accretion models, the deuterium burning sequence defines a stellar 
birthline in the HR diagram.  Previous calculations find similar results.
Stahler (1983, 1988; see also \cite{pal90}, 1993) first identified the 
birthline for spherical accretion and showed that this locus provides 
a good upper envelope for observations of young stars in nearby molecular 
clouds.  Mercer-Smith \etal (1984) published the first HR diagram track 
for a disc-accreting protostar using a code similar in spirit to the 
Eggleton code.  Hartmann \etal (1997) later derived a birthline for disc 
accretion from semi-analytic calculations.  Our deuterium burning
sequence lies close to Stahler's (1983, 1988) birthline for log
$T_{\rm eff} \gtrsim$ 3.55 and falls $\delta$ log $L \approx$ 0.1--0.2
below the birthline for log $T_{\rm eff} \lesssim$ 3.55.  The displacement 
reflects differences in the starting conditions and outer boundary
condition.  The birthlines converge at large $T_{\rm eff}$, because the
boundary conditions become less important as the stellar luminosity
increases (see also \cite{pal90}, 1993).  

The location of the stellar birthline in our models is sensitive to the 
adopted deuterium abundance.  This behavior follows from the explicit
dependence of the stellar luminosity on the rate of deuterium burning
(see, for example, \cite{sta88}; \cite{har97}).  Our deuterium burning 
sequence shifts by $\delta$ log $L \approx$ $-$0.13 for a factor of two 
reduction in $\rm X_D$.  Our birthline then roughly coincides with that 
of Hartmann \etal (1997) for $\rm X_D = 1.75 \times 10^{-5}$.  

With only six class I sources in our HR diagram, it is difficult 
for the data to favor convincingly any theoretical calculation.  
The 2$\sigma$ error bars are consistent with all of the tracks, 
even without considering uncertainties in the model input parameters.  
The data lie closer to the disc-accretion birthline of Hartmann \etal
(1997) and our deuterium burning sequence than either the birthline for 
spherical accretion or the accretion track of Mercer-Smith \etal (1984).
In all cases,  changing model input parameters -- such as the
deuterium abundance -- would allow a better match between data and
the models.  More rigorous comparisons thus await observational 
estimates of unknown quantities such as the deuterium abundance 
and a larger sample of class I sources with reliable spectral types 
and luminosities.
	
We conclude this section with several points about the comparison of 
observations with model tracks.  First, the distribution of class I 
sources about {\it any} birthline should be uniform, if the range in 
initial conditions is small.  This dispersion should be comparable to 
the observational errors.  In our case, 5 out of 6 class I sources
fall 1--2$\sigma$ below the birthline for spherical accretion 
(\cite{sta83}, 1988) and the Mercer-Smith \etal (1984) accretion track.  
The data are somewhat more consistent with our deuterium burning 
sequence and the Hartmann \etal (1997) birthline for an accretion rate 
of $10^{-5} ~ \rm M_\odot\,{\rm yr}^{-1}$.  A larger sample of class I
sources should distinguish between models.

Second, it is important to compare the {\it stellar} component
of the protostellar luminosity with the predictions of model tracks.
The observed class I luminosities are the {\it total} luminosity and
have not been corrected for the unknown amount of accretion luminosity.
In most of the CTTS shown in Figure 8, the accretion luminosity is a 
small fraction -- 10\% to 30\% -- of the stellar luminosity plotted 
in the figure (see \cite{har91}, 1995; \cite{gul98}).  However,
the accretion luminosity is roughly comparable to the stellar
luminosity in the continuum + emission sources and is $\sim$ 100
times the stellar luminosity in FU Ori systems like L1551 IRS5
(\cite{har91}; \cite{har96}).
We expect a small accretion contribution for class I sources with
optical spectra similar to most CTTS, although the accretion 
luminosity in L1489 IRS may be large (see also \cite{gr96b}).  
The arrow in Fig. 8 indicates the displacement in log $T_{\rm eff}$ 
for a $\delta$ log $L$ = $-$0.1 change in the bolometric luminosity.
This change moves the data closer to both the birthline for
spherical accretion and our deuterium sequence and illustrates 
the difficulty in comparing models with current data.  

Third, knowledge of the deuterium abundance, and to a lesser extent
the lithium abundance, is also necessary to compare observations with 
model predictions.  The mean deuterium abundance for stars in a
molecular cloud sets the location of the birthline in the HR diagram;
any star-to-star scatter in the abundance spreads the birthline 
vertically in the HR diagram.  To our knowledge, the deuterium 
abundance has not been measured in any pre--main-sequence star.  
Recent measurements indicate a factor of 3--5 scatter in the 
lithium abundance among nearby molecular clouds (e.g., \cite{kin93}; 
\cite{lee94}; \cite{dun96}) that could be due actual abundance 
differences between stars (see \cite{lee94}) or differences in 
the analysis procedures (see \cite{dun96} and references therein).
Observations of older open clusters may also indicate considerable
star-to-star differences in the rate of lithium depletion among stars 
with the same mass (e.g., \cite{ran97} and references therein).
Similar spreads in the deuterium burning rate, due perhaps to
star-to-star variations in accretion rate, further complicates 
the comparison of observations with model tracks.
The sample of protostars is not currently large enough to worry
about abundance variations, but the uncertainties will become
more important as sample sizes increase.

\section{DISCUSSION and SUMMARY}

In the previous sections, we have described the first optical
spectroscopic survey of class I, embedded sources in a single
molecular cloud.  We supplemented these data with high quality
optical spectra of a representative sample of older and optically
brighter T Tauri stars.  The combined set of spectra show that
the optical spectra of class I sources qualitatively resemble 
the optical spectra of T Tauri stars.  Our analysis further
reveals common physical properties and substantial differences 
between class I sources and T Tauri stars, as summarized below.

Our data provide the first indication that the distribution of 
stellar spectral types among class I sources may not be very
different from that of WTTS and CTTS.  Of the five class I sources
with strong optical continua, one (L1489 IRS) is a continuum +
emission source, three are M-type stars, and another (04264+2433)
may have an M-type central star.  To the best of our knowledge, {\it 
these are the first low mass protostars with measured optical spectral 
types.}
This sample is too small for a meaningful comparison with 
the distribution of spectral types among more evolved 
pre--main-sequence stars in the cloud.  We note, however, that 
the median spectral type for WTTS and CTTS is K7-M0 and that the 
frequency of continuum + emission sources is $\sim$ 5\%--10\% (KH95).

Published observations indicate other similarities between 
class I sources and older pre--main-sequence stars in Taurus-Auriga.
First, class I sources have the same intrinsic near-IR colors 
as do CTTS.  Whitney \etal (1997) show that the observed near-IR 
colors of class I sources can be modeled as a CTTS surrounded by an 
infalling envelope with an optical extinction, $A_V \approx$ 30--60 mag.
This analysis leads to the conclusion that the radiation from
the star and inner disc of a class I source is similar to that
of a T Tauri star (see also \cite{gr96b}; Calvet \etal 1997 reach
a different conclusion).
Second, the bolometric luminosity distributions of class I sources,
CTTS, and WTTS are indistinguishable (KH95).  All three groups of
pre--main-sequence stars have median luminosities of $L_{\rm b}
\approx$ 0.5--0.8 \lsun.  This unexpected result is supported
by the positions of class I sources in the HR diagram.  Our data 
show that class I sources have luminosities and effective
temperatures very similar to those of CTTS and WTTS in the cloud.  
These conclusions are surprising, because a class I source should 
have a larger luminosity once it has accreted nearly all of its 
final mass, and this luminosity should decline with time as the 
star approaches the main sequence (see, for example, \cite{sta83}, 1988;
\cite{pal93}; \cite{har97}; Fig. 8).  The current sample, however, 
is too small to test stellar models in detail. The errors in 
luminosity and effective temperature are also too large.  
Observations with the next generation of ground-based telescopes 
will undoubtedly expand the sample, reduce the errors, and 
provide better tests of protostellar accretion theory.
 
One feature that distinguishes class I sources is their strong 
forbidden-line emission.  As a group, class I sources are much more
likely to have forbidden-line emission than CTTS or WTTS (Fig. 7).
This result confirms previous conclusions from imaging data (e.g.,
\cite{gom97}) and indicates that class I sources are more likely to
drive outflows than CTTS or WTTS (see also \cite{bon96}; \cite{mor92},
1994).  The equivalent widths of the forbidden lines are also larger 
in class I sources than in CTTS or WTTS.  Although some large 
equivalent widths may be due to very weak optical continua, 
the [S~II] equivalent widths in HH31 IRS2 -- a class I source 
with a prominent TiO absorption band -- are larger than observed 
in {\it any} CTTS in our sample (see Tables 1--2).  Deeper optical 
spectra of our sample and other class I sources would clarify this point.

Our sample is not large enough to test whether class I sources also 
have more prominent {\it permitted} emission lines than CTTS.  
The median H$\alpha$ equivalent width for class I sources,
$\sim$ 90 \AA, is much larger than the median equivalent width 
for CTTS, $\sim$ 30--40 \AA.  This difference is roughly what we
expect if class I sources have larger continuum veiling than CTTS
(\cite{cas96}; \cite{gre97}) and if the H$\alpha$
equivalent width correlates with veiling (\cite{har95} and references
therein).  However, the frequency of He~I $\lambda\lambda$5876, 6678 
emission among class I sources is roughly comparable to that among CTTS.
We measure a He~I emission frequency of 50\% among 6 class I sources
with reasonable signal-to-noise at 6000 \AA, 
57\% among 14 flat-spectrum sources, and 65\% among 46 class II sources.
For both emission lines, the class I sample is probably biased 
against small equivalent widths, because class I sources without
emission lines are probably fainter than sources with emission lines.
A deeper survey with a larger telescope could enlarge the sample of
class I sources with high quality optical spectra.  These data would 
provide a good test for differences in the distribution of H$\alpha$ 
equivalent widths between class I sources and CTTS.

These results fit into the general picture of Taurus-Auriga class I 
sources developed in KH95 and in Kenyon \etal (1990).  In this picture,
class I sources are envelopes of gas and dust falling into the 
central star-disc system at rates of a few $\times~10^{-6}~\msunyr$
(see also \cite{ada87}; \cite{ke93a}, 1993b; \cite{whi97}).  
Bell \& Lin (1994) show that the stable accretion rate through the disc
onto the central star is either very low -- $\lesssim$ a few 
$\times~10^{-7}~\msunyr$ -- or very high -- $\gtrsim$ a few 
$\times~10^{-5}~\msunyr$ -- compared to the infall rate.  The
disc spends most of its time in the low accretion rate state; 
the disc mass then  slowly increases with time until it reaches 
a critical level and evolves to the high accretion rate state.  
This model explains the low observed luminosities of nearly all
class I sources as well as the occasional high luminosity of a 
source such as L1551 IRS5.  

Models with time-dependent disc accretion also qualitatively 
account for the evolution of forbidden and permitted emission
lines in pre--main-sequence stars.  We expect the time-averaged
accretion rate through the disc to decline as the envelope 
disperses.  If the H$\alpha$ and other permitted emission lines
of class I sources form in the accretion region of the inner disc
as in CTTS, then the median H$\alpha$ equivalent width should decline
as a pre--main sequence star evolves from a class I source into 
a CTTS and then into a WTTS.  Most models for jet formation link 
the mass loss rate in the jet to the mass accretion rate through
the disc (see, 
for example, \cite{cab90}; \cite{naj94}; \cite{sh94a}, 1994b), 
so we expect forbidden emission to decline as well.
Explaining the observations of emission line equivalent widths 
with a quantitative model of a dispersing envelope and evolving disc, 
however, is currently beyond our reach.

Finally, our results further demonstrate the advantages of optical
spectra.  Recent surveys of larger samples of class I sources 
using near-IR spectroscopy have yielded only two spectral types each
in Taurus-Auriga (\cite{cas96}; \cite{gr96b}) and $\rho$ Oph 
(\cite{gr96b}, 1997).
Casali \& Eiroa (1996; see also \cite{cas92}; \cite{gr96b}, 1997) 
conclude that continuum emission from dust in a circumstellar disc 
or envelope veils photospheric absorption features on near-IR spectra
of class I sources.  Preliminary results further suggest that this 
veiling is larger in class I sources than in CTTS or WTTS 
(\cite{cas96}; \cite{gr96b}, 1997). Dust emission is much weaker 
relative to a normal stellar photosphere at shorter wavelengths,
$\lesssim 1~\mu$m, so optical spectra may yet provide the best 
measure of spectral types in class I sources.

\vskip 6ex
We thank the staffs of the MMT, Palomar, and Whipple Observatories
for assistance with our observations.  Fred Chaffee kindly acquired
several spectra of the class II sources listed in Table 1.
Susan Tokarz reduced the FAST 
spectra and graciously assisted with the reduction of the MMT and 
Palomar spectra.  We also thank M. Geller, M. G\'omez, C. Lada, 
A. Mahdavi, and B. Whitney for advice and comments.  The suggestions
of an anonymous referee improved our presentation.
Observations at the Palomar Observatory were made as part of a
continuing collaborative agreement between Palomar Observatory and
the Jet Propulsion Laboratory.
Portions of this research were supported by the National Aeronautics
and Space Administration through grant NAGW-2919 and by the Space
Telescope Science Institute through grant GO-06132.01-94A.
C.A.T. thanks the Royal Society and the Hungarian Academy of Sciences 
for an exchange fellowship during the majority of his contribution 
to this project.

\vfill
\eject

\vfill
\eject

\epsfxsize=8.5in
\epsffile{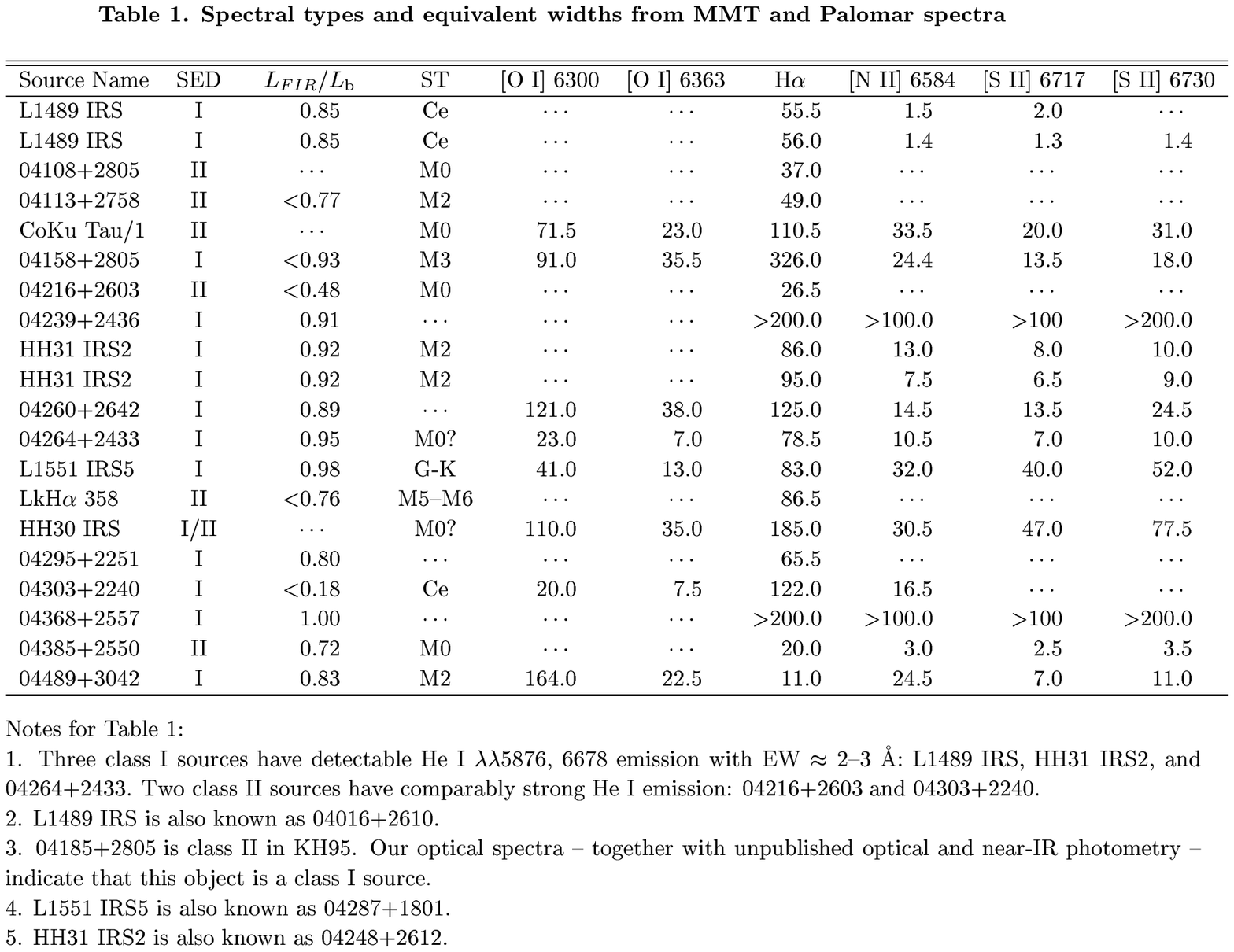}

\epsfxsize=8.5in
\epsffile{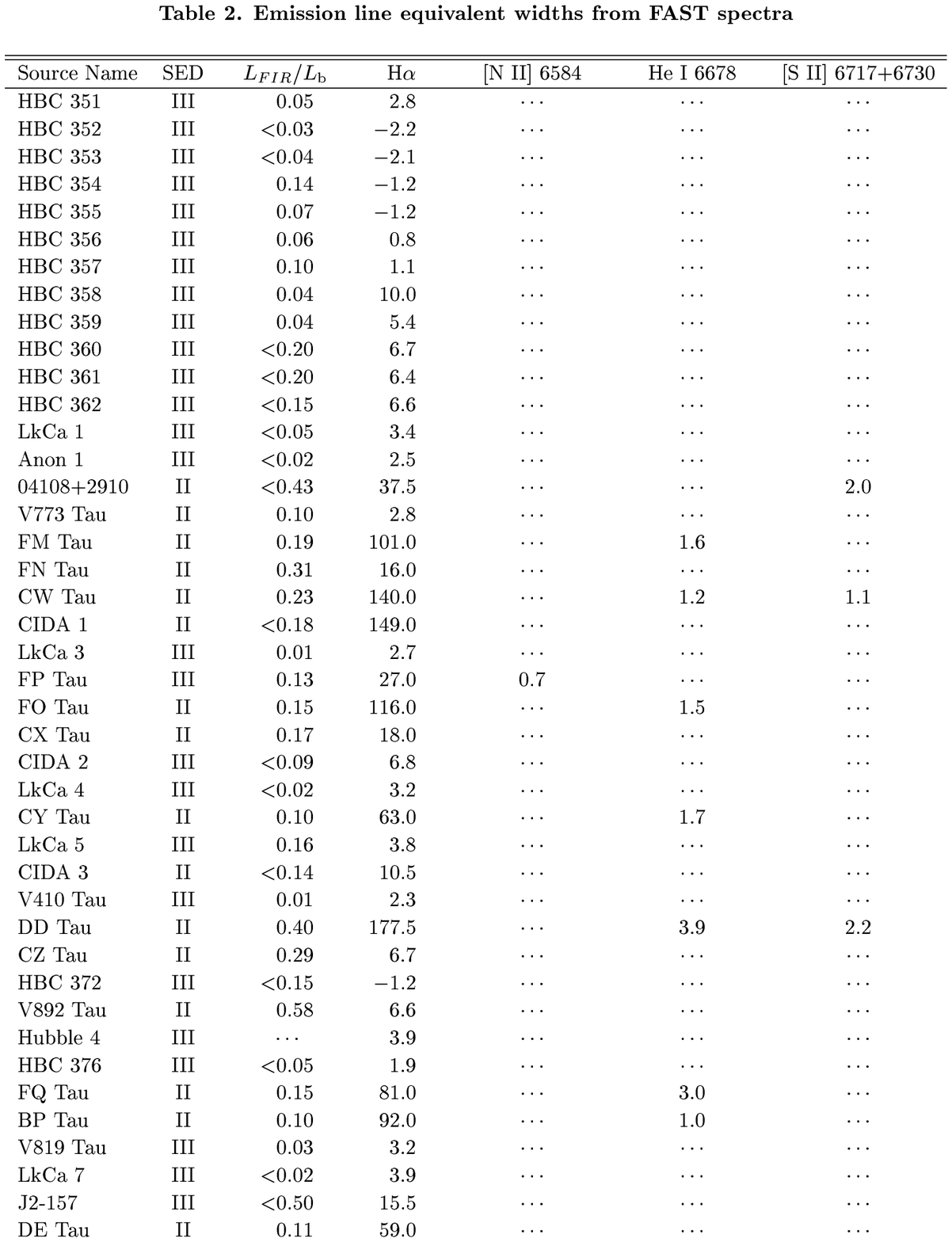}

\epsfxsize=8.5in
\epsffile{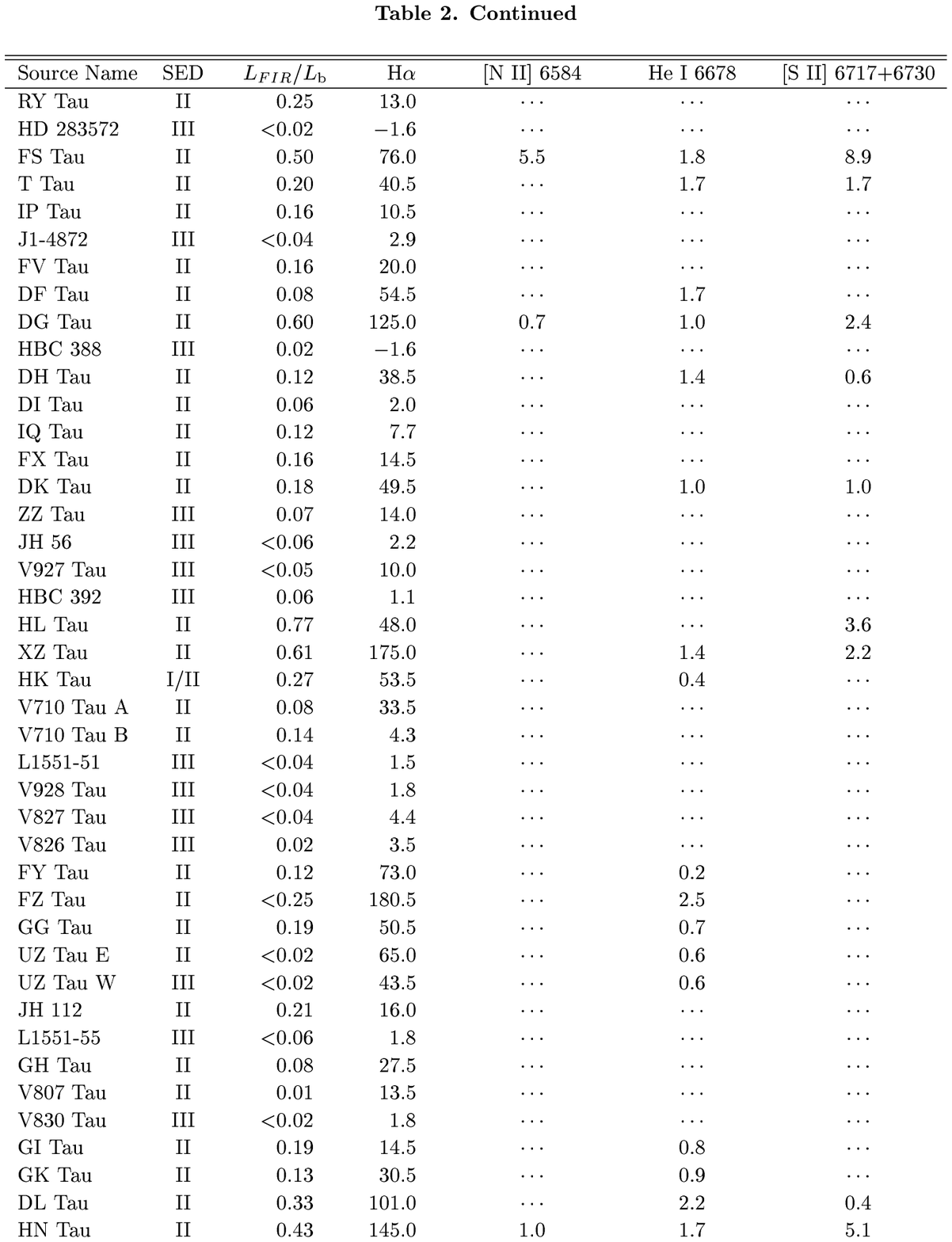}

\epsfxsize=8.5in
\epsffile{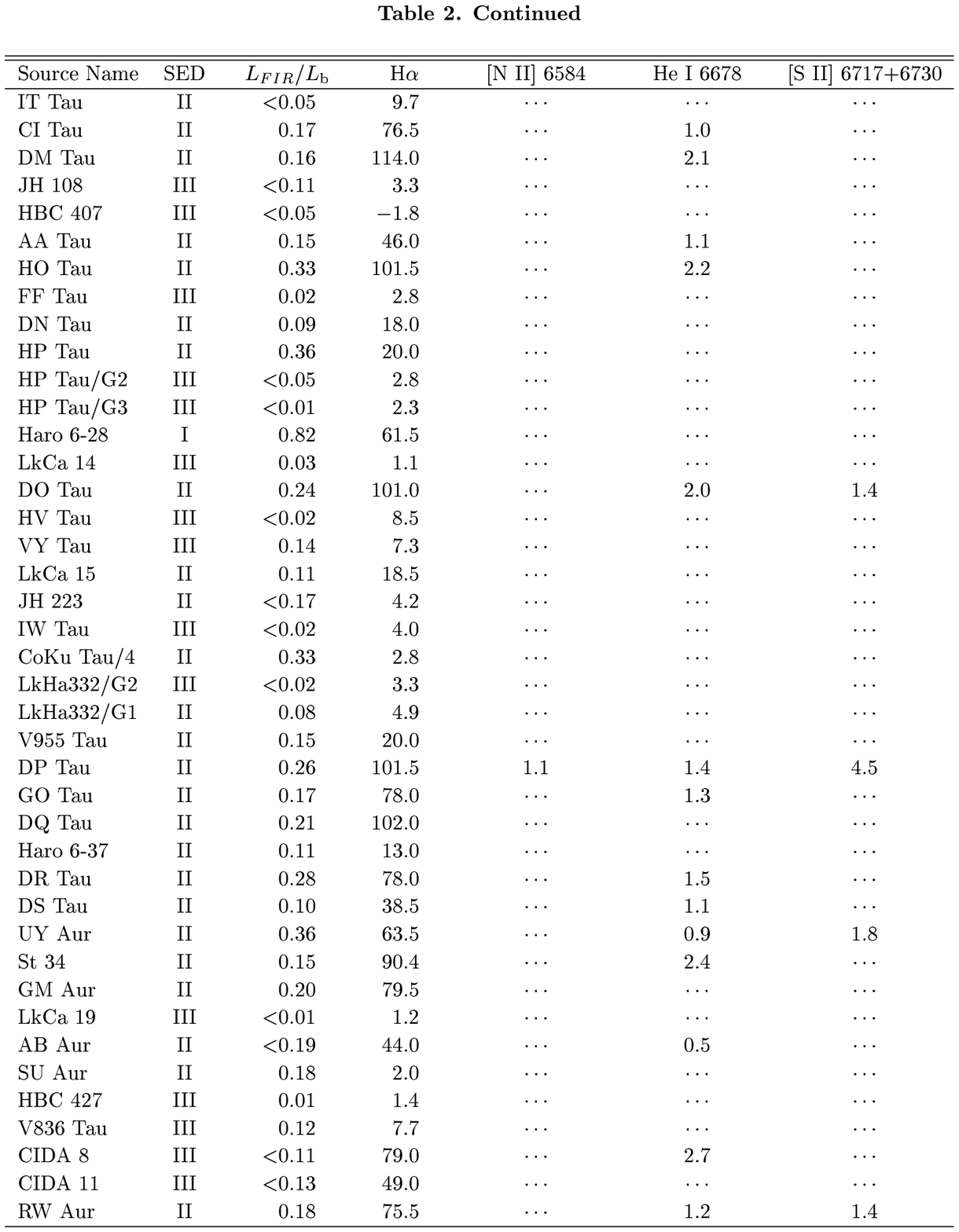}

\centerline{\bf FIGURE CAPTIONS}
\vskip 4ex

\singlespace

\figcaption[Kenyon.fig1.ps]
{Optical spectra of T Tauri stars. The degree of excitation
increases clockwise from the weak emission line T Tauri star, 
LkCa 3, through the classical T Tauri stars, BP Tau and DP Tau,
to the continuum + emission star DG Tau.}

\figcaption[Kenyon.fig2.ps]
{Spatially resolved spectra of Taurus pre--main-sequence stars.
The left panels plot H$\alpha$ spectra for two class I sources,
L1489 IRS (04016+2610) and HH31 IRS2 (04248+2612).  
The right panels plot [S~II] spectra for L1527 IRS and HH30 IRS.
All sources are very extended compared
to the 1\arcsec--2\arcsec~extent of an unresolved point source.}

\figcaption[Kenyon.fig3.ps]
{Optical spectra of four class I sources in Taurus-Auriga.
All class I sources have strong emission from H$\alpha$, [O~I],
and [S~II]; some also have He I and Ca II triplet emission.
Two systems -- HH31 IRS2 and 04489+3042 -- have the deep TiO
absorption bands characteristic of M-type stars. 
A telluric absorption feature at $\lambda$7650 is present in
the spectra of L1489 IRS and 04264+2433; this feature is blended
with a TiO band in HH31 IRS2 and 04489+3042.}

\figcaption[Kenyon.fig4.ps]
{Optical spectra of two Taurus-Auriga class I sources.
One class I source, 04158+2805, has deep TiO absorption bands
and strong H$\alpha$, [O~I], and [S~II] emission lines.
The spectrum of HH30 IRS is similar to other Herbig-Haro objects,
with very intense emission from H$\alpha$, [O~I], [N~II], and
[S~II]. A telluric absorption feature at $\lambda$7650 is present in
the spectrum of HH30 IRS; this feature is blended with a TiO band 
in 04158+2805.}

\figcaption[Kenyon.fig5.ps]
{Optical spectra of four class I sources in Taurus-Auriga.
These objects have negligible continuum emission and very strong
emission from H$\alpha$, [O~I], [N~II], and [S~II].
The emission line equivalent widths are comparable to those 
observed in Herbig-Haro objects.}

\figcaption[Kenyon.fig6.ps]
{TiO indices for field dwarfs and Taurus-Auriga 
pre--main-sequence stars.  Several class I sources (diamonds) and 
heavily reddened T Tauri stars (plusses) have TiO band strengths
comparable to those observed in T Tauri stars with negligible
optical veiling (light filled triangles) and normal main-sequence stars
(filled circles).}

\figcaption[Kenyon.fig7.ps]
{Frequency of [S~II] emission among Taurus-Auriga 
pre--main-sequence stars.  The [S~II] emission frequency
increases with the ratio of far-IR to bolometric luminosity,
$L_{FIR}/L_{\rm b}$.}

\figcaption[Kenyon.fig8.ps]
{HR diagram for Taurus-Auriga pre--main-sequence stars.
The filled circles are T Tauri stars from Kenyon \& Hartmann (1995).
The crosses are class I sources from this study.  The error bar
in the upper right corner shows the typical uncertainty in
luminosity and effective temperature for a class I source.
Uncertainties for T Tauri stars are $\sim$ 33\%--50\% of the 
uncertainty for class I sources.  The arrow indicates the change 
in log $L$ and log $\rm T_e$ if a continuous source of emission 
veils the optical continuum.  The dashed lines plot isochrones 
for the pre--main-sequence tracks of
D'Antona \& Mazzitelli (1994) at times of $10^5$, $10^6$, and
$10^7$ yr from top to bottom.  The dot-dashed line is the stellar
birthline from Stahler (1988).  The solid lines are pre--main-sequence
tracks for stars accreting from discs at $10^{-5} \msunyr$ 
(thicker line) and $10^{-6} \msunyr$ (thinner line).  }

\vfill
\eject

\pagestyle{empty}

\epsfxsize=8in
\epsffile{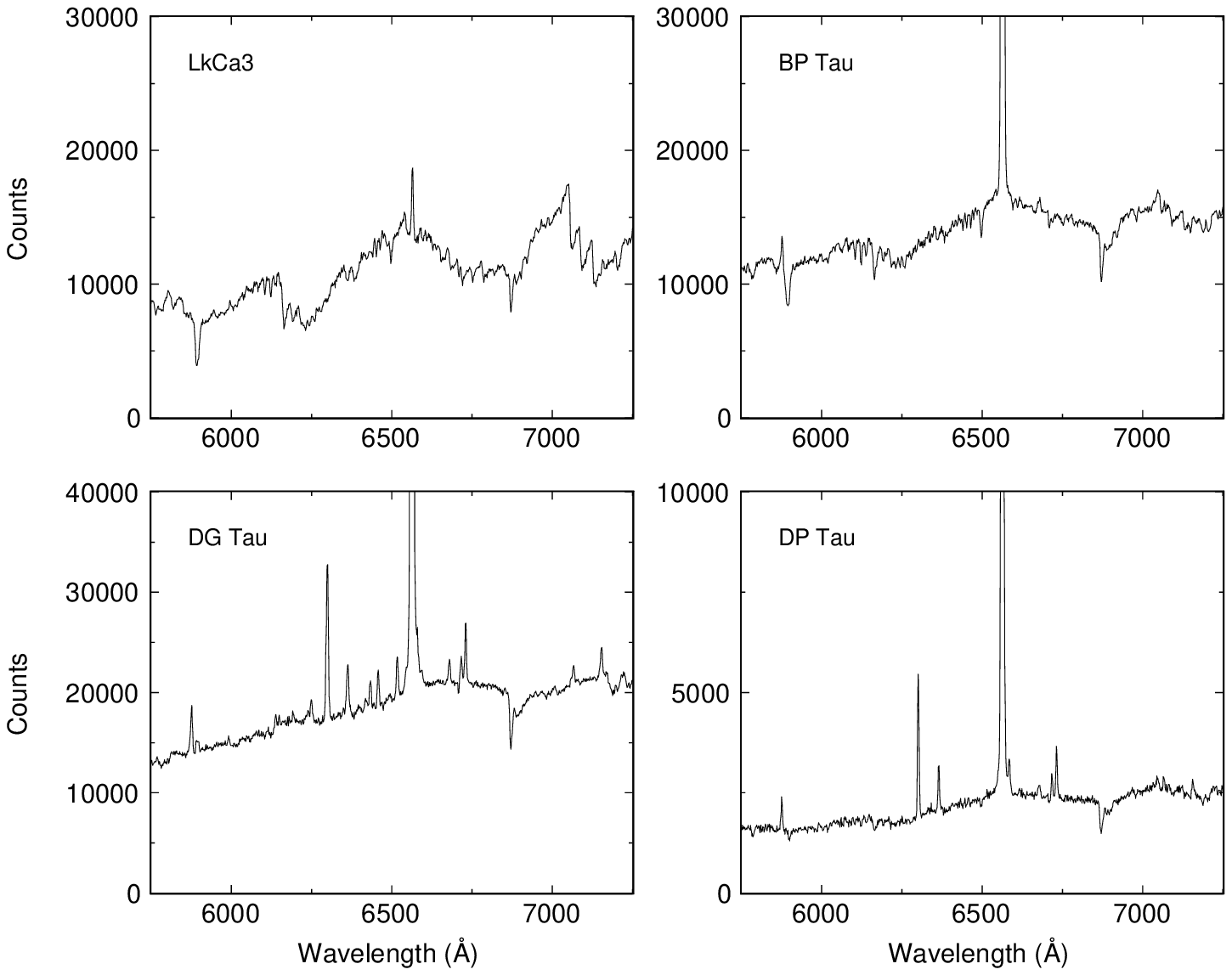}
\vskip 5ex
\centerline{\bf {\large Figure 1}}

\epsfxsize=8in
\epsffile{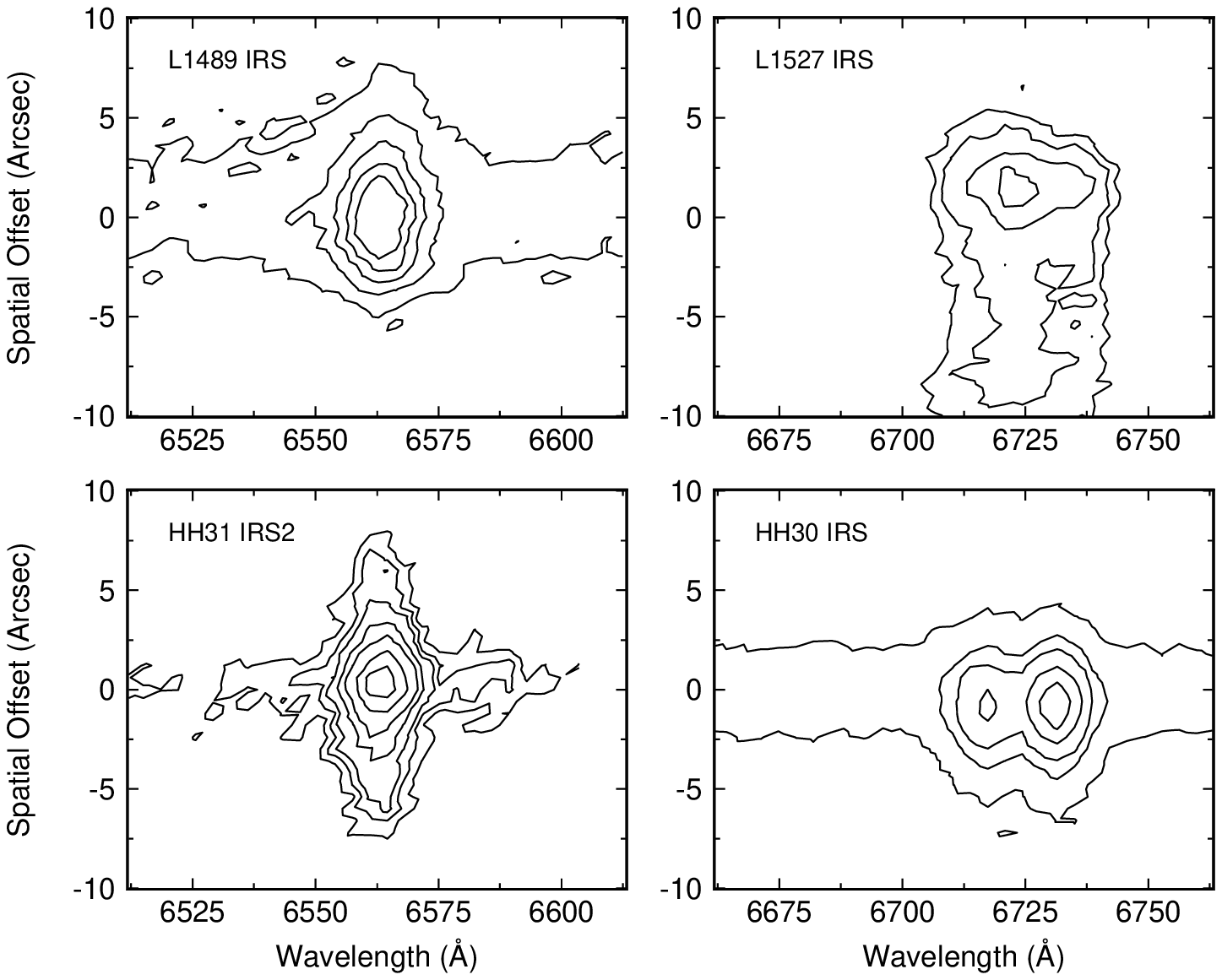}
\vskip 5ex
\centerline{\bf {\large Figure 2}}

\epsfxsize=8in
\epsffile{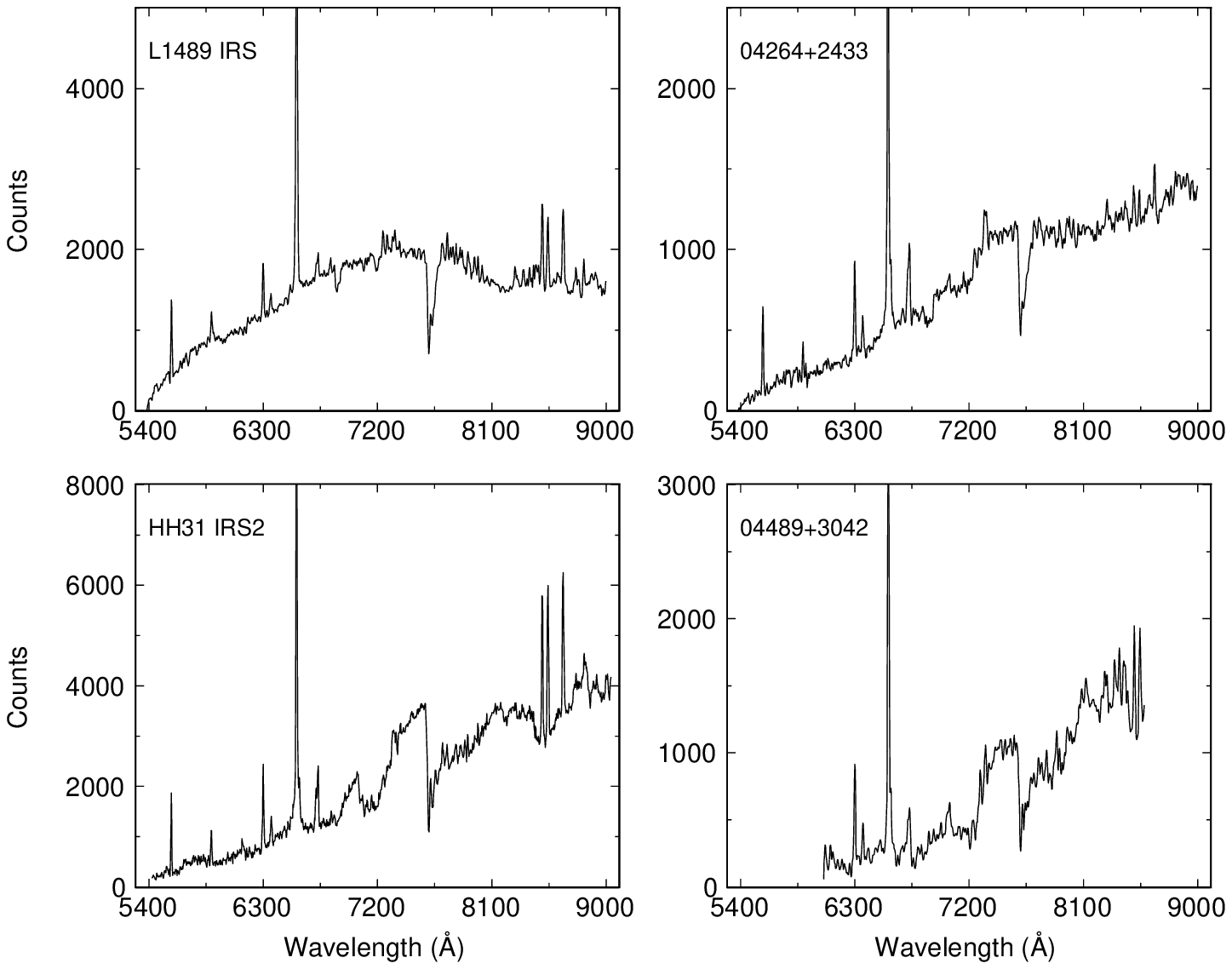}
\vskip 5ex
\centerline{\bf {\large Figure 3}}

\epsfxsize=8in
\epsffile{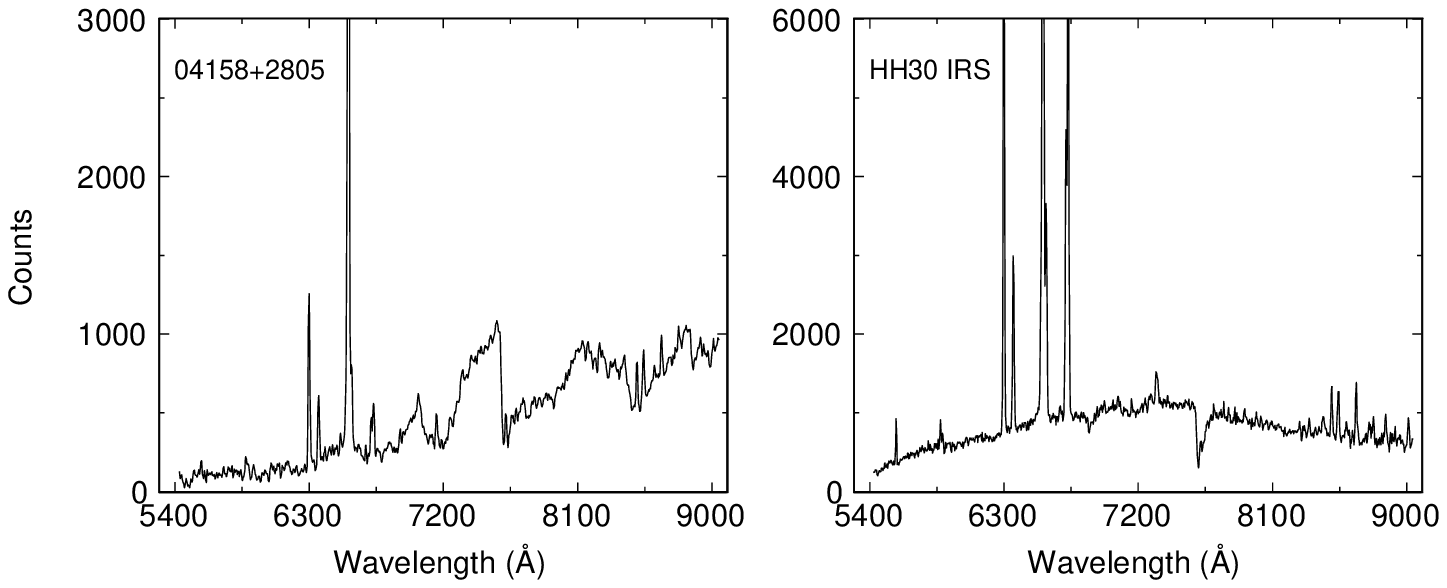}
\vskip 5ex
\centerline{\bf {\large Figure 4}}

\epsfxsize=8in
\epsffile{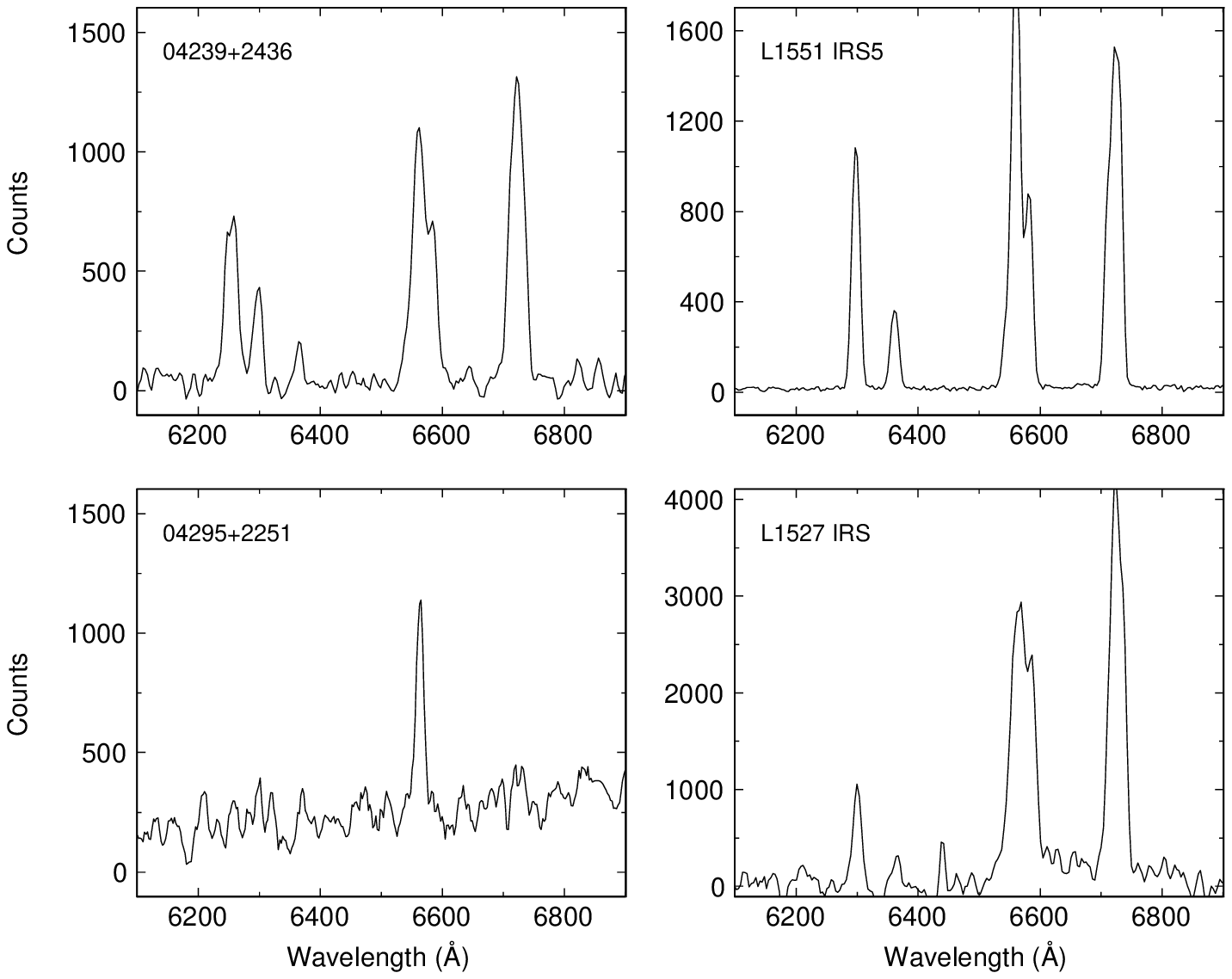}
\vskip 5ex
\centerline{\bf {\large Figure 5}}

\epsfxsize=8in
\epsffile{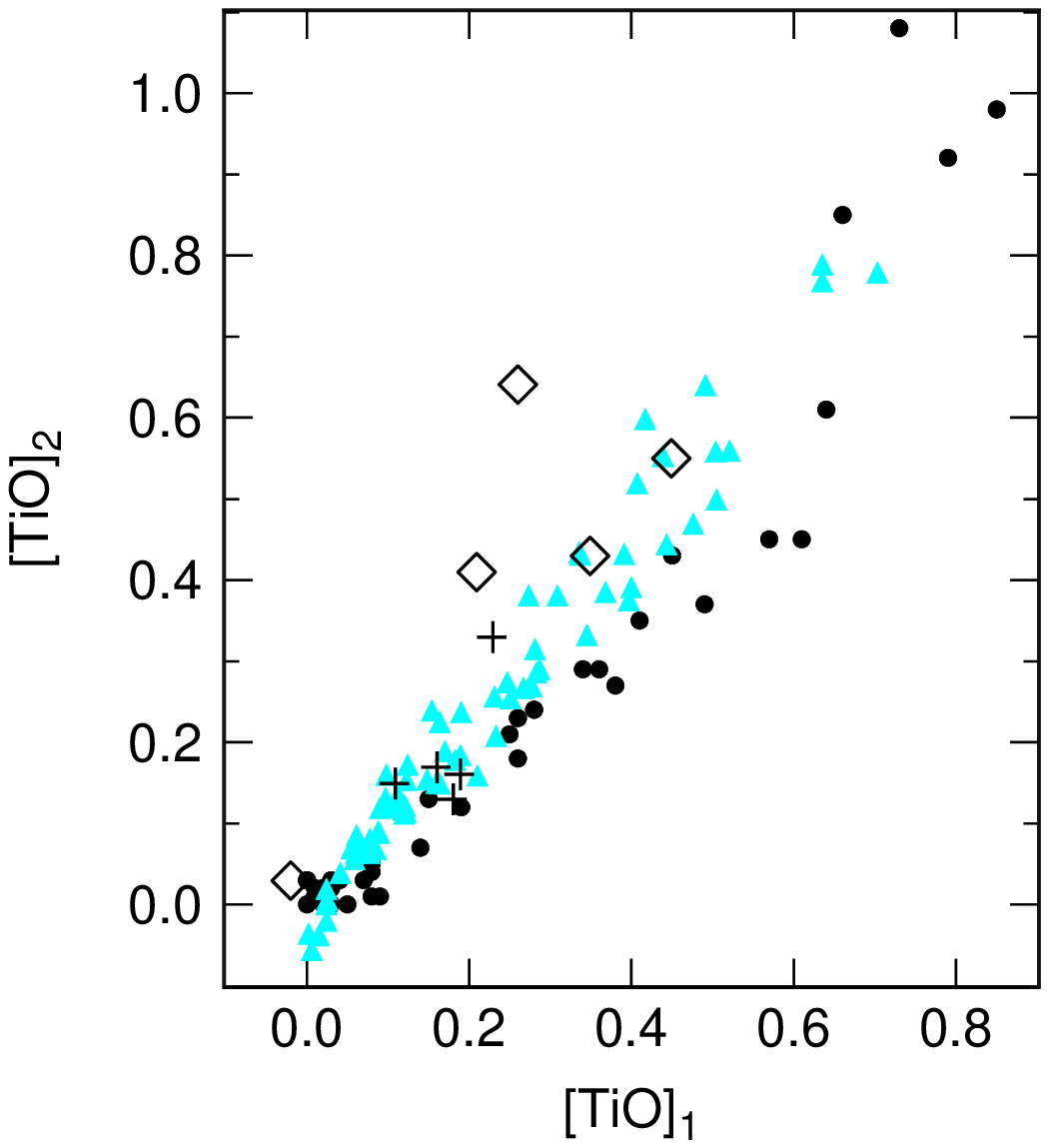}
\vskip 5ex
\centerline{\bf {\large Figure 6}}

\epsfxsize=7.5in
\epsfxsize=8in
\epsffile{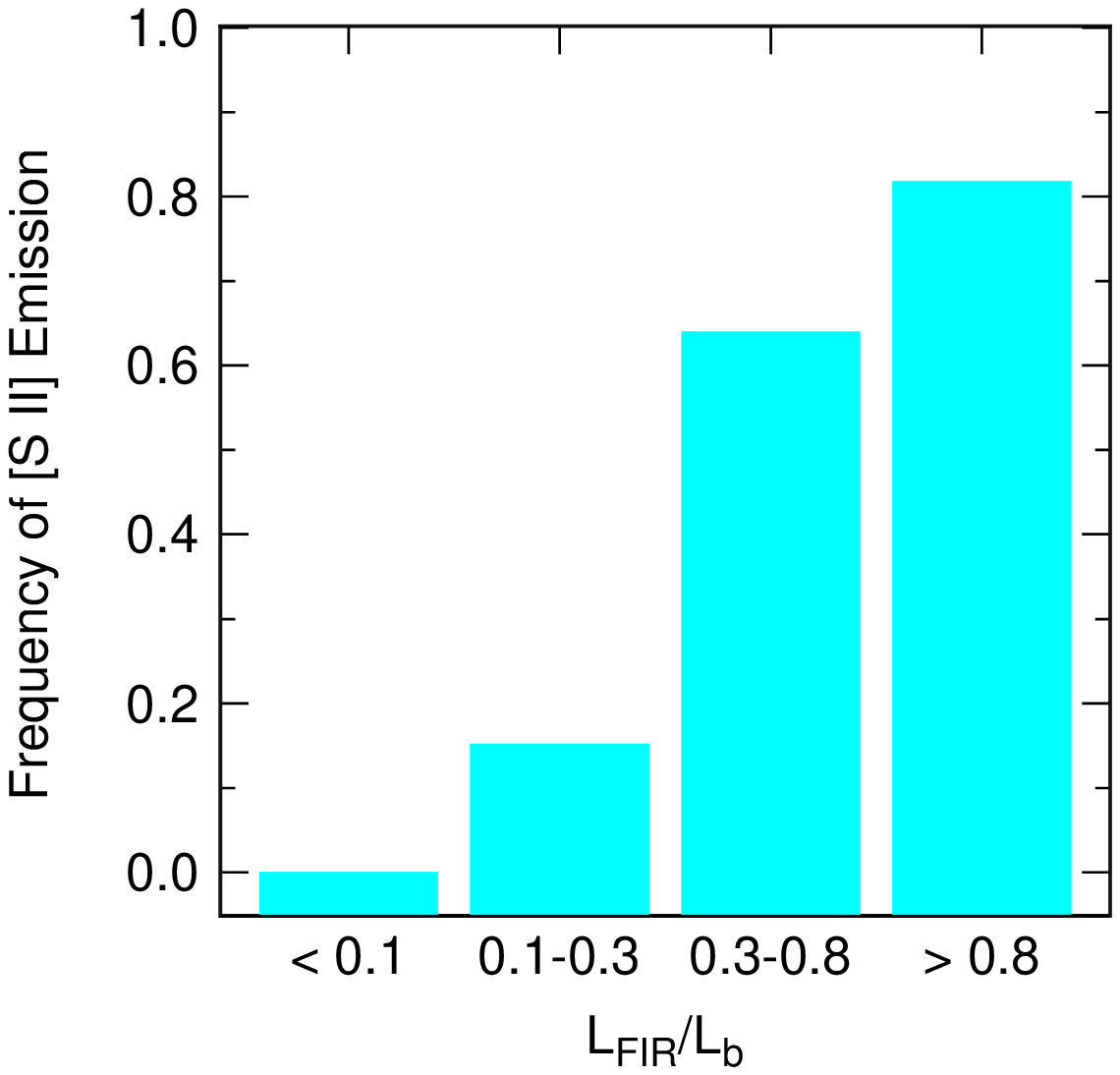}
\vskip 5ex
\centerline{\bf {\large Figure 7}}

\epsfxsize=7.5in
\epsffile{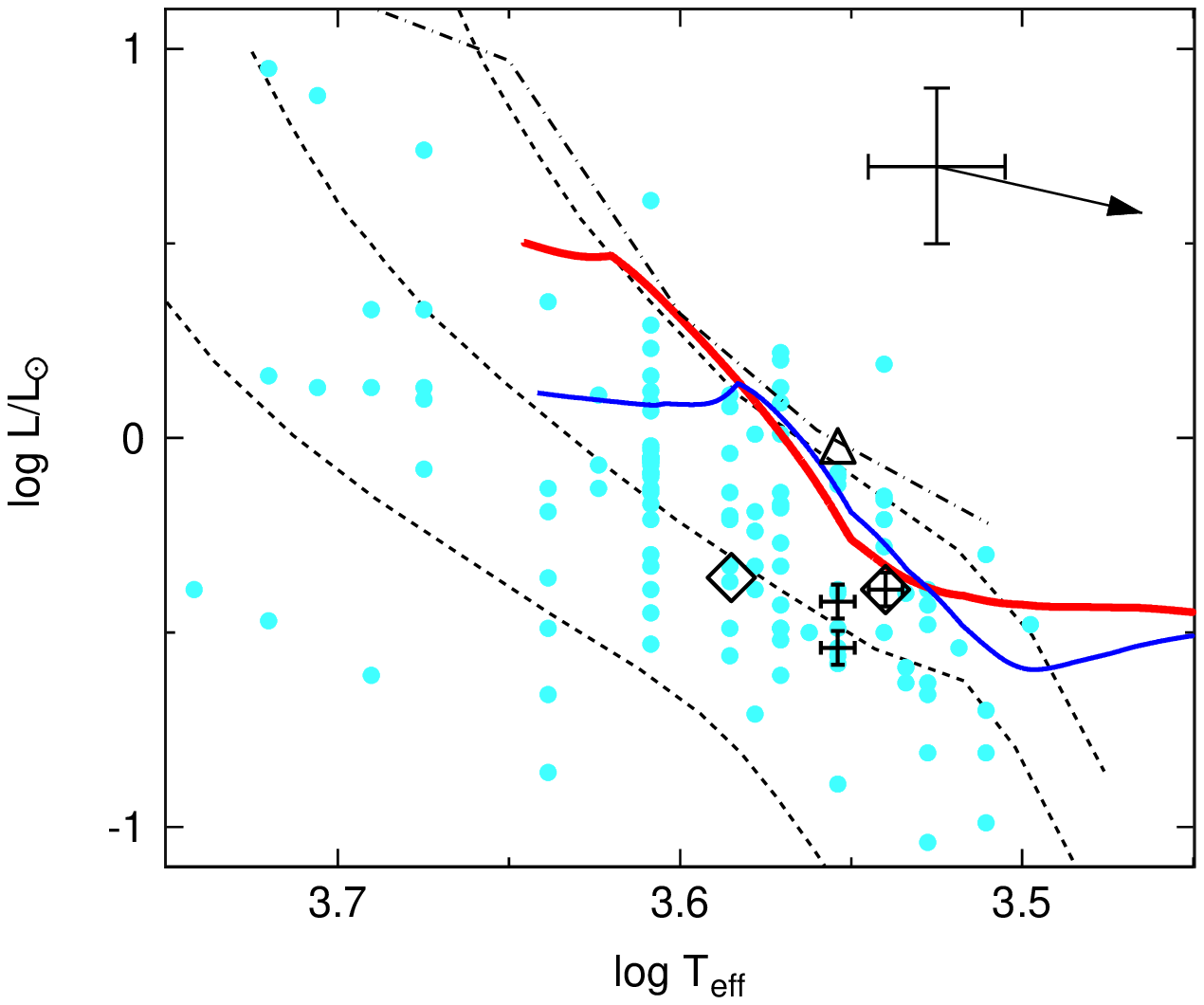}
\vskip 5ex
\centerline{\bf {\large Figure 8}}

\end{document}